\documentclass{elsart.new}
\usepackage{natbib}
\usepackage{epsfig}
\begin{document}
\begin{frontmatter}
\title{Particle acceleration in ultra-relativistic parallel shock waves}
\author[IMP,MPIfR]{A. Meli\thanksref{Mailaddr}} and
\author[IMP]{J.J. Quenby}

\address[IMP]{Astrophysics Group, Blackett Laboratory, Imperial College
of Science, Technology and Medicine, Prince Consort Rd. SW7 2BW,
London, UK.}
\address[MPIfR]{Visiting Max Planck Institut fuer Radioastronomie, 
Auf dem Huegel 69 53121, Bonn, Germany.}
\thanks[Mailaddr]{Corresponding author: a.meli@ic.ac.uk}

\begin{abstract}
Monte-Carlo computations for highly relativistic parallel shock
particle acceleration are presented for upstream flow gamma factors,
$\Gamma=(1-V_{1}^{2}/c^{2})^{-0.5}$ with values between 5 and $10^{3}$.
The results show that the spectral shape at the shock depends on whether or not
the particle scattering is small angle with $\delta \theta <
\Gamma^{-1}$ or large angle, which is possible if $\lambda > 2r_{g}
\Gamma^{2}$ where $\lambda$ is the scattering mean free path along
the field line and $r_{g}$ the gyroradius, these quantities being measured in the 
plasma flow frame. The large angle scattering case exhibits distinctive 
structure superimposed on the basic power-law spectrum, largely absent in the pitch angle case. 
Also, both cases yield an
acceleration rate faster than estimated by the conventional,
non-relativistic expression, $t_{acc}=[c/(V_{1}-V_{2})]
[\lambda_{1}/V_{1}+\lambda_{2}/V_{2}]$ where '1' and '2' refer to upstream and downstream 
and $\lambda$ is the mean free path. A $\Gamma^{2}$
energy enhancement factor in the first shock crossing cycle and a significant energy 
multiplication in the subsequent shock cycles are also observed. The results may
be important for our understanding of the production of very high energy cosmic rays
and the high energy neutrino and gamma-ray output from Gamma Ray
Bursts (GRB) and Active Galactic Nuclei (AGN).
\end{abstract}

\end{frontmatter}


\section{Introduction}

There are three distinct astrophysical situations where the bulk plasma
flow is extremely relativistic with Lorentz factors
$\Gamma=(1-V^{2}/c^{2})^{-1/2} \geq 10$.
These are some Active Galactic Nuclei (AGN) jet and hot spot sites, Gamma Ray Burst (GRB) fireballs and 
pulsar polar winds. Within these flows relativistic shocks appear and diffusive
particle acceleration takes place. In each case, there is evidence for
energetic particle acceleration to high $\Gamma$ factors but the upper
limit to the possible energy attained
becomes less certain with increasing bulk flow velocity. For AGN jets,
$\Gamma \sim 10$ plausibly results in the known 
$10^{19}$ eV cosmic rays (Quenby and Lieu, 1989) via diffusive
shock acceleration at the termination shocks. However, the more direct synchrotron/self-Compton
modeling only reveals 10 GeV gamma-rays (e.g. Maraschi et al., 1992).
For GRB fireballs, $\Gamma\sim 100-10^{3}$ 
appears to eventually produce gamma-rays at least up to 100 MeV (Fenimore et al., 1993) but 
at such flow speeds, it is not clear whether repeated shock crossings are possible and the
predictions of $10^{19}$ eV neutrinos (Vietri, 1998), which would be direct
evidence of a diffusive like process have yet to be verified.
Similarly, pulsar winds could result in a $\Gamma \sim 10^{4}$ flow
encountering the nebula envelope (Hoshino et al., 1992), but there is no evidence
for more than TeV acceleration anywhere in the system. However Lucek and Bell (1994) show that combined
motion across a near perpendicular shock with a non-uniform field and in the
electromagnetic field of the nebula may produce much higher energies.\\
Past work has suggested
that if the shocks are close to parallel so that an ${\bf E}=0$ frame
can be defined, diffusive shock acceleration is in principle possible and provided large angle 
scattering occurs in the plasma frame, each shock crossing cycle yields a $\sim \Gamma^{2}$ energy increase
(Quenby and Lieu, 1989). This result has been exploited by Vietri (1995,1998) modeling GRB fireball particle
shock acceleration. It is the purpose of this paper to present simulation
results on particle  acceleration with $\Gamma$ values extended up to $\sim$ $10^{3}$.
Here we concentrate on the  parallel shock configurations ($\psi=0^{o}$).
Our results, measured in the shock frame, will be compared  with the standard spectral prediction of
non-relativistic theory, where the
momentum spectrum is $n(p)\propto p^{-\alpha}$ with $\alpha=(r+2)/(r-1)$, for shock
compression ratio $r=V_{1}/V_{2}$, independent of scattering details, and $\alpha=2$ for
strong unmodified shocks with $r=4$. Analytically it has been shown that over a limited momentum
range, the non-relativistic time scale for acceleration is
$T=[c/(V_{1}-V_{2})][\lambda_{1}/V_{1}+\lambda_{2}/V_{2}]$ where, '1'
and '2' refer to upstream and downstream and $\lambda$ is the scattering
mean free path. The shock jump solutions of Ballard and Heavens (1991) and the 
review of Kirk and Duffy (1999) show that, a range of shock frame
$r$ values largely limited between 3 and 4, typically arise for oblique, relativistic shocks, 
whereas $r$=4 is the extreme relativistic hydrodynamic limit. Hence for simplicity we adopt 
$r$=4 but, in the companion paper dealing with oblique shocks we demonstrate the insensitivity of our chief 
conclusions on the exact value. For our parameters, $T=(8/3)(nr_{g,1}/c)$  where, $\lambda=nr_{g}$ and $r_{g}$ 
is Larmor radius.\\
The first relativistic correction noticed to the non-relativistic theory,
was the spectral flattening seen in parallel shocks (Kirk and Schneider, 1987a,b), followed by 
the discovery of a speed-up in the acceleration rate (Quenby and Lieu, 1989; Ellison et al., 1990). 
These results have subsequently been extended to non-parallel and non-linear sub-luminal
shocks (Lieu, et al., 1994 and Bednarz and Ostrowski, 1996 who use $T=r_{g}/c$ as the unit of 
acceleration time), where there is large angle scattering
or pitch angle diffusion. 
Most recently, much interest has been focused on differences occurring
between small angle and large angle scattering models for the
test particle-turbulence interactions in the fluid frame. Gallant and
Achterberg (1999) suggest that for a field model consisting
either of randomly orientated uniform field cells or a uniform field, the
scattering is limited to $\delta \theta < 1/\Gamma$ in the
upstream fluid frame. Generally, such a model yields a test particle
differential spectral slope $\sim -2.2$ (Baring, 1999). Furthermore, on this
model, Gallant and Achterberg (1999) provided an analytical demonstration that in the high $\Gamma$ 
limit, only the first crossing cycle exhibited a $\Gamma^{2}$ energy increase,
while in all other crossings the energy is only doubled and hence
one may expect the relativistic speed-up to be limited to this crossing. 
The computations of Baring (1999) also showed a much reduced
energy enhancement in subsequent cycles for the small angle scattering case. 
Achterberg et al. (2001) extend the conclusions of Gallant and Achterberg (1999) to a wider variety of 
upstream field models while, Kirk et al. (2000) employ an eigenfunction method to deal with the
anisotropies in diffusion when the scattering depends on the angle with the shock normal.
A recent general review of particle acceleration and relativistic shocks is given by Kirk and Duffy (1999). 
However, the detailed contrast between the effects of various particle scattering models and the 
behaviour of these models at $\Gamma$ factors $\sim 10^{3}$ remain to be elucidated, both for parallel 
and quasi-perpendicular shocks. Hence this paper
starts a series of investigations in the ultra-relativistic limit with a consideration of parallel shocks. 
In section 2 we present the numerical method used, section 3 presents the results 
while in section 4 conclusions and consequences are given. A further paper seeks to 
extend this work to non-parallel shocks (Meli and Quenby, 2002b).


\section{Numerical Method}

The actual purpose of these simulations is to find a solution to the
particle transport equation for highly relativistic parallel flow
velocities. The appropriate time independent Boltzmann equation is given
by the following as we assume steady state,

\begin{equation}
\Gamma(V+\upsilon\mu )\frac{\partial f}{\partial x}=\frac{\partial f}{\partial t}\arrowvert_{c}
\end{equation}
 
where $\Gamma$ is the Lorentz factor of the fluid frame, $V$ is the fluid
velocity, $\upsilon$ the velocity of the particle,  $\mu$ the cosine of
the particle's pitch angle and $\frac{\partial f}{\partial t}|_{c}$
the collision operator. The reference frames we use
during the simulations are the upstream and downstream fluid
frames and the shock frame. Pitch angle is measured in the local fluid
frame, while $x$ is in the shock rest frame and the model assumes 
variability in only one spatial dimension. In order for the above equation to be
solved by a simple Monte Carlo technique, we assume that the collisions represent scattering in pitch 
angle with no cross-field diffusion, the scattering is elastic in the fluid frame and 
a phase averaged distribution function may be employed. We will discuss these assumptions in more detail
in due course. The collision operator may be characterized by two types of
scattering that the particles suffer, large angle and small angle scattering,
which are respectively represented by the two following equations:

\begin{equation}
\frac{\partial f}{\partial t}|_{c}=\frac{<f>_{\mu}-f}{\tau}
\end{equation}

where $\tau$ is the mean time between collisions, related to the mean free path by the relation 
$\lambda=\upsilon\tau$ and,

\begin{equation}
\frac{\partial f }{\partial t}|_{c}=\frac{\partial}{\partial
\mu}(D_{\mu\mu}\frac{\partial f}
{\partial \mu})
\end{equation}

where $D_{\mu\mu}$ is the pitch angle diffusion coefficient.
As mentioned previously, Gallant and Achterberg (1999) have demonstrated that small angle 
scattering with $\delta \theta < 1/\Gamma$ applies with $\theta$ measured in the upstream fluid 
frame for scattering in a uniform  field or a randomly orientated set of uniform field cells. This 
arises because particles attempting to penetrate upstream from the shock are swept back into 
the shock before they can move far in pitch. To establish the conditions for large angle scattering, 
we demand the particle penetrating upstream can move over about a Larmor radius, $\sim r_{g}$, 
on encountering the scattering centre while still upstream. A suitable scattering centre would 
be a rotational discontinuity or a much enhanced field region. Let $\upsilon $ be the particle velocity in 
the shock frame, directed perpendicular to the shock in the upstream direction, the most favourable 
case for re-entry upstream.  If $\upsilon \sim c$,  transforming to the upstream frame via 
$\upsilon_{1}=(\upsilon-V_{1})/(1-\frac{\upsilon V_{1}}{c^{2}})$ and allowing 
$V_{1}=c-\delta \sim c$ yields $\upsilon \sim c$. Now in the upstream frame, the shock velocity 
is $-c+\delta$, if the positive spatial coordinate, $x$, is towards the shock. If $\lambda_{1}$ 
is the mean free path for scattering (encountering a large angle scatterer) in the
upstream frame, the average time to reach this scatter centre after leaving the shock is 
$\Delta t_{1}=\lambda_{1}/c$. Meanwhile, the shock moves in the upstream frame a distance 
$(c-\delta)\Delta t_{1}$. Hence we require $r_{g,1}<\lambda_{1}-(c-\delta)\Delta t_{1}$ to allow 
space for the upstream large angle scatter to occur.
If $\Gamma_{1}=[1-(V_{1}^{2}/c^{2})]^{-0.5}$, we find $r_{g}<\lambda_{1}/2\Gamma_{1}^{2}$ 
as the condition for large angle scattering. \\
The most favourable site to find high $\Gamma$ large angle scattering is in GRB which may arise 
from collapse resulting in a rapidly rotating black-hole torus system (van Putten, 2001).
Rotation allows a $1/r$ field and we start from a torus of about  one Schwarzchild radius, depending 
on spin, of a $20 M/M_{\circ}$ hypernova collapse and the presence of a field near the torus of about
$B_{\circ}=10^{14}$ gauss. Although Brainerd (1992) 
shows that  for typical GRB parameters and a cosmological jet model, cyclotron and 
synchrotron radiation are only  possible for fields below about $10^{10}$ gauss,
his arguments apply only to the regions from which emission emerges. 
Lieu et al. (1989) are among authors pointing out the 
extra quantum mechanical processes arising at $\Gamma B > 10^{14}$, 
thus suggesting that the maximum possible field strength, which could arise in equipartition
with the burst binding energy, $\sim 10 ^{18}$  gauss, will rapidly reduce 
in strength as discussed by Lerche and Schramm (1977).    
Blobs, rather than a steady 'wind' arise in some models (e.g. Blackman et al., 1996).
Meszaros et al. (1993) suggest a comoving fireball size during
the main acceleration phase $r_{b,1}\sim 10^{14}\rightarrow10^{16}$ cm and
a mean fireball field just due to equipartition with internal shocks
of $B_{a,1}\sim10^{2}$ gauss. The suffix '1' is used for comoving quantities and this 
suffix is omitted for fixed frame quantities. The transverse 
comoving frame 'bullet' fields which are likely to be effective for
our scattering are estimated by $B_{s}=B_{\circ}r_{\circ}/r$
where $B_{1}=B/\Gamma$ and $r_{1}=\Gamma r$. Let $\lambda_{a,1}$ be
the parallel mean free path in the ambient or general shocked field
outside the 'bullets' with, $\lambda_{a,1}=nr_{g,a,1}$ where, $r_{g,a,1}$
is Larmor radius and $n>10$ as suggested by detailed computations in the 
turbulent interplanetary medium (Moussas et al., 1992 and references therein). 
Diffusive shock acceleration can hardly be expected
unless the mean free path is significantly smaller than the dimensions of the region.
Also $r_{g}\propto\gamma$, the particle Lorentz factor.
Hence we may write $\lambda_{a,1}=min[r_{b,1}/10,n\rho_{ 1,a}]$.
Then inserting this expression in our 
large angle scatter limit requires either $B_{a,1}/B_{s,1}<n/2\Gamma^{2}$ independent of $\gamma$ or, 

\begin{equation}
p_{1}(eV)<\frac{300B_{s,b,1}r_{b,1}}{10\times2\Gamma^{2}}(gauss,cm)
\end{equation}

where, the particle momentum $p_{1} $ and field $B_{s,b,1}$ are measured near the fireball boundary.
Taking the extreme $\Gamma=10^{3}$, with $n$ = 40, the first inequality is marginally satisfied 
if $r_{b,1}\leq10^{14}$ cm while the second inequality is satisfied at a moving frame
momentum of $<10^{16}$ eV or rest frame momentum of $<10^{19}$ eV.
Thus, it is possible to envisage GRB models with $\Gamma<10^{3}$ where large angle scattering is occurring,  
especially if a larger torus or torus field is allowed.\\
Here we consider one-dimensional parallel shocks ($\psi=0^{\circ}$, where
$\psi$ is the angle between the shock normal and the magnetic field),
either because that is the field configuration, or because turbulence removes
'reflection' at the interface even though the magnetic field does not lie along the $x$-axis.
A test particle approximation is used for simplicity to allow a step by step approach 
to understanding the full problem, though eventually non-linear effects must be included.
Relativistic particles of initial $\gamma\sim (\Gamma+10)$ are injected upstream towards the  
shock and are allowed to scatter in the respective fluid frames with their basic motion
described by the guiding centre approximation.
One justification is we note that Bell (1987) and Jones and Ellison (1991) have shown that
'thin' sub-shocks appear even in the non-linear regime, so at some energy above the plasma 
value, the accelerated particles may be dynamically unimportant while they recross the discontinuity.
Another way of arriving at the test particle regime is if 
particles are injected well above the plasma particle energy, remaining dynamically unimportant and
thus requiring the seed particles to have been already pre-accelerated.
In AGN jets, traveling shocks superimposed on the relativistic flows accelerated by pressure 
at the jet base, could provide the seed for the terminal 'hot spot' acceleration.
Such seed particles appear in the neutron star binary merger scenario for GRB
(Narayan et al., 1992), which includes the presence of pre-accelerated particles
before any terminal shock acceleration phase, 
possibly due to explosive reconnection. For isolated pulsars, we note
that there is strong observational evidence (e.g. PSR1913+16, PSR2127+11C) 
for the continuous injection of relativistic particles into the surrounding medium, over their lifetime.
While this plasma will probably be predominantly in the form of
$e^{+}e^{-}$ pairs, created in the pulsar magnetosphere, it has been argued that pulsar winds must
also contain ions in order to account for the electric currents
in the Crab Nebula (Hoshino et al., 1992; Gallant and Arons, 1994). 
These authors employ a wave acceleration process at the shocked termination of the flow, to 
provide a non-thermal pair spectrum to yield the X-ray and gamma-ray output. However their input
distribution function is also not mono-energetic and therefore containing 'seed particles' above
the bulk flow energy. The supposed nature of the slot/gap electric field acceleration of the pair 
plasma is unlikely to give a mono-energetic output.\\ 
A relativistic transformation is performed from the local plasma frames to the 
shock frame to check for shock crossings. In this parallel shock case, there is no reflection allowed 
and on crossing  the computation is continued in the other plasma frame. We change units so that 
$m\approx c=1$. The particle is made to leave the system if it 'escapes' 
far downstream at the spatial boundary, or if it reaches a well defined maximum energy 
$E_{max}=E_o 10^{14}$, for computational convenience, even though other physics 
describing particle escape or energy loss would probably need to be taken into account 
in realistic situations.  The downstream spatial boundary required
can be estimated from the solution of the convection-diffusion
equation in a non-relativistic, large-angle scattering 
approximation in the downstream plasma which gives the 
chance of return to the shock, $exp(-V_{2}r_{b}/D)$, yielding a probability of return of $2\cdot10^{-3}$
if $r_{b}=8\lambda_{\|}$. In fact, runs are performed with different spatial boundaries to investigate
the effect of the size of the acceleration region on the spectrum, as well as to find a region where
the spectrum is size independent. In the small-angle scatter case, the inherent anisotropy due to the
high downstream sweeping effect may greatly modify this analytical estimate.\\
The particles ($\sim$$10^{5}$) of a weight ($w_p$) equal to 1.0, are injected far upstream and they
are allowed to move towards the shock, along the way colliding with the scattering centers. Provided multiple
scattering between the upstream and downstream regions of the shock can occur
and they gain energy in each crossing, diffusive shock acceleration is simulated. The
compression ratio ($r$) is kept at the value of 4 for simplicity.

A splitting technique is applied, similar to Bednarz and Ostrowski's (1998) in order to obtain
statistical accuracy over a wide range of particle Lorentz factors. 
The splitting technique helps to avoid the consequence that in highly relativistic 
flow environments, only a few high energy accelerated particles, 
which remain in the acceleration process, dominate the recorded distribution function, 
thereby limiting the statistical accuracy to an energy range of only two or three decades
above injection energy $E_o$. 
By introducing more particles, but of corresponding lower weight when some energy 
well above injection is attained, a greater sampling of the possible regions 
within the accessible phase space of the model is achieved.\\
The mean free path $\lambda$ for particle scattering is calculated in the respective fluid frames
and if it is assumed that it is dependent on the particle's energy as given by
the formula, $\lambda=\lambda_o p cos\theta$, where $\theta$ is the
particle's pitch angle, so as to allow for a particle 'current' and phase space weighting,
the probability that a particle will move a distance $x$ along the  field
lines at pitch angle $\theta$ before a scattering is given by the expression,
$ p(x)\propto exp(-x/\lambda)$. At the scattering centers the energy
(momentum) of the particle is kept constant and only the direction of the velocity vector ${\bf
\upsilon}$ is randomized, using a computational random number generator. The downstream 
mean free path is taken as a factor 4 less than upstream 
while the absolute value is arbitrary since the model boundaries are specified 
in units of the mean free path.\\
To model the large angle scatter case, the new pitch angle is chosen at random 
(phase angle is neglected in this guiding centre approximation but will need to be chosen
to specify the scattering).
For the picture of pitch angle diffusion case, we assume that the tip of a
particle's momentum vector undergoes randomly a small change $\delta\theta$ in its direction
on the surface  of a sphere and within a small range of polar angle (after a  small increment of time) 
(Ellison et al., 1990). If the particle had an initial pitch angle $\theta$, 
we calculate its new pitch angle  $\theta'$ by the simple trigonometric formula,

\begin{equation}
cos\theta'=cos\theta\sqrt{1-sin^2\delta\theta}+sin\delta\theta\sqrt{1-cos^2\theta}cos\phi
\end{equation}

where, $\phi\in(0,2\pi)$ is the azimuth angle with respect to
the original momentum direction and must be chosen at random in this case. 
For the case of highly relativistic flows, we limit particles pitch angle diffusion to angles 
chosen at random, up to an angle $\sim 1/\Gamma$, where $\Gamma$ is the upstream gamma, 
measured in the shock frame (Gallant and Achterberg, 1999) in order to be consistent with these 
authors' model for upstream re-entry but some runs are performed with 
the scattering angle $\sim 0.1/\Gamma$ to correspond to the later work of Achterberg et al. (2001) 
and Protheroe et al. (2002) -private communication.\\
A consequence of this limitation in upstream scatter for the Gallant and Achterberg (1999) model
follows if we calculate the ratio of energies crossing down-upstream 
measured in the respective fluid frames as $\Gamma(1+\beta_{r} \mu_{\rightarrow u}^{'})$ where
$\beta_{r}$ is the relative velocity of the two streams as a fraction of $c$ and $\mu_{\rightarrow u}^{'}$
is cosine pitch angle for down to upstream crossing in the downstream
frame. This angle only needs to exceed $1/4 \rightarrow1/ 3$  for kinematic reasons.
Hence the energy gain can still be $\sim \Gamma$ in contrast to the gain ratio for the up to down
transition which is $\Gamma(1-\beta_{r}\mu_{\rightarrow d})$ and where, $\mu_{\rightarrow d}$
the upstream frame cosine pitch angle on up to down transmission is nearly
unity due to the limitation on the scattering permissible.\\
This prediction of limited energy gain on all crossings subsequent to the first cycle, will 
be investigated in our simulations, as it is important to establish  
the dependence of the model results on whether or not large angle or pitch angle diffusion 
operate in highly relativistic flows.


\section{Results}

First, test runs to verify the validity of the codes which have been performed in the
mildly relativistic limit (e.g. $V_{1}=0.1c-0.6c$) tend to excellent agreement with the
non-relativistic theory, giving smooth spectra shapes and spectral indexes close to -2.2.
In addition, similar runs show that as the flow becomes more relativistic, the
spectrum flattens, which is in agreement with the analytical and numerical work
of Kirk and Schneider (1987a,b). We also compare our results with similar
studies for $\Gamma\sim 5$ (Baring, 1999) and find agreement concerning
the spectral index and the spectral shape for both large angle scattering and pitch
angle diffusion mechanisms.\\
As we have already mentioned a compression ratio ($r$) of 4 is used for simplicity and also to
allow immediate comparison with non-relativistic unmodified strong 
shock results, but our preliminary code runs show
that the qualitative trend of the results is insensitive to the exact
compression ratio value, due to the mildly-relativistic nature of the downstream plasma.
Initial Lorentz factors of the flow investigated are 5, 50, 200, 500
and 990. 

All results are given at the shock, in the shock frame.
The particle distribution function is obtained in a particular energy and $\mu$ space cell by recording 
the passage of each simulation particle crossing the cell, 
weighted by the time taken in cell crossing.\\
In figure 1 we show the  ratio of the computational time constant to the
non-relativistic analytical acceleration time constant, as defined in the 'Introduction' section,
calculated as a function of the upstream Lorentz factor $\Gamma$ flow for large and small 
angle scattering. Note that we are effectively measuring 'speed-up' of the acceleration time in units 
of a few times $r_{g}/c$. 

\begin{figure}[t]
\begin{center}
\epsfig{file=plot.lg.time, width=5.0cm}
\epsfig{file=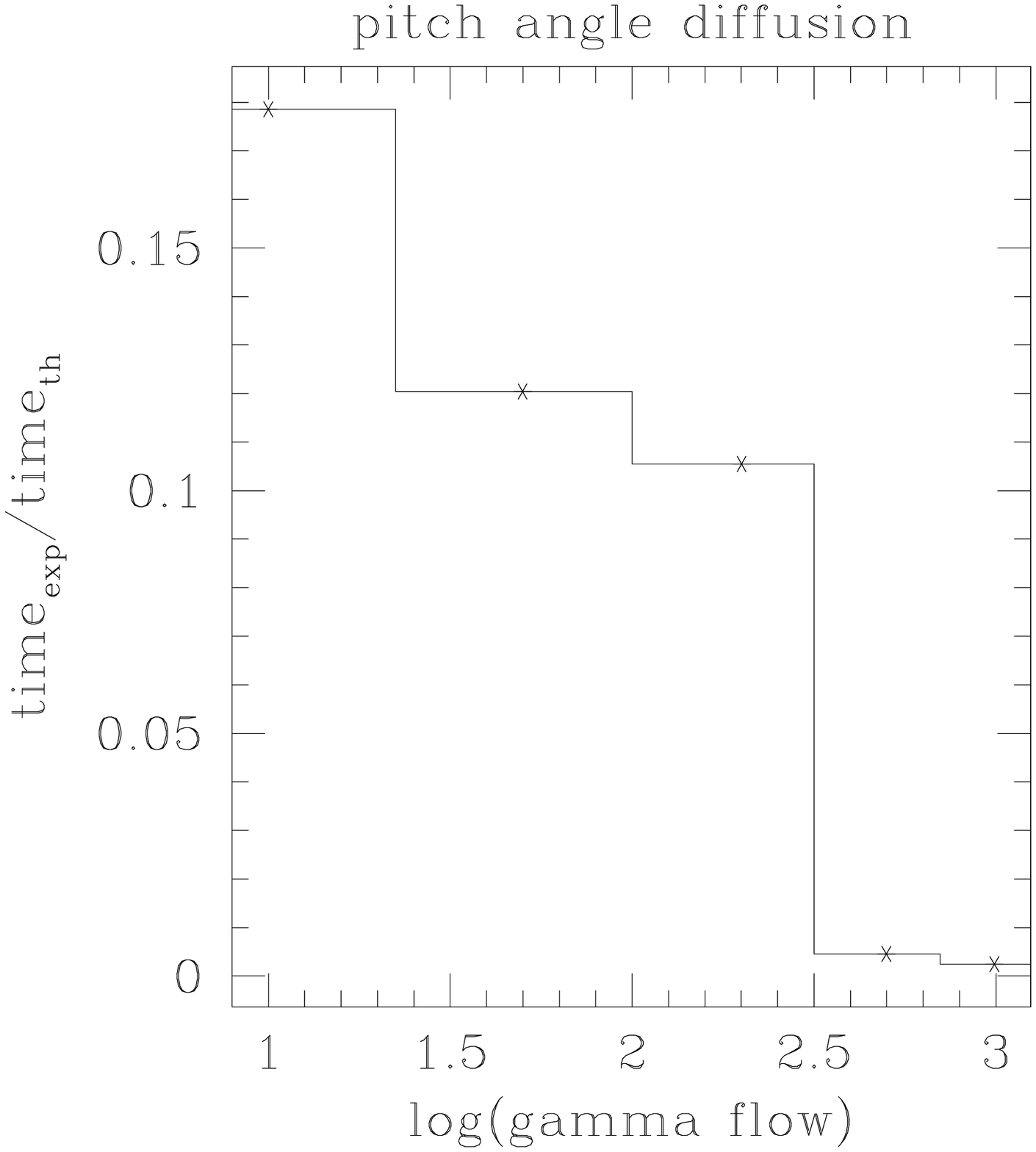, width=5.0cm}
\caption{The ratio of the computational time to the non-relativistic
analytical constant (measured in the shock frame)  versus the logarithm of the Lorentz 
$\Gamma$ flow, for large angle scattering (left) and for pitch angle diffusion (right).}
\end{center}
\end{figure}

We obtain the time constant $t_{acc}$ from the computation
by recording the energy increment for each complete cycle, up-down-up and the time taken 
for the cycle at a particular
energy level, $\Gamma$ and finding the mean increments, $\Delta \Gamma$ and $\Delta t$. 
Then $t_{acc}/\Delta t=\Gamma/\Delta\Gamma$.  
We find a 'speed-up' of about a factor of 5 for large angle scattering and a considerable
speed-up of a factor $\sim$ 20 for pitch angle diffusion. In this latter case, the parallel mean free path
is estimated for the analytical formula from the time to multiple scatter in pitch angle through
$\pi/2$. That is $\lambda_{\|}=(\pi^{2}/4)(c\lambda/\delta \theta^{2})$ where 
$\lambda\propto \gamma$ is the step length between collisions yielding $\delta \theta$. 

\begin{figure}[h!]
\begin{center}
\epsfig{file=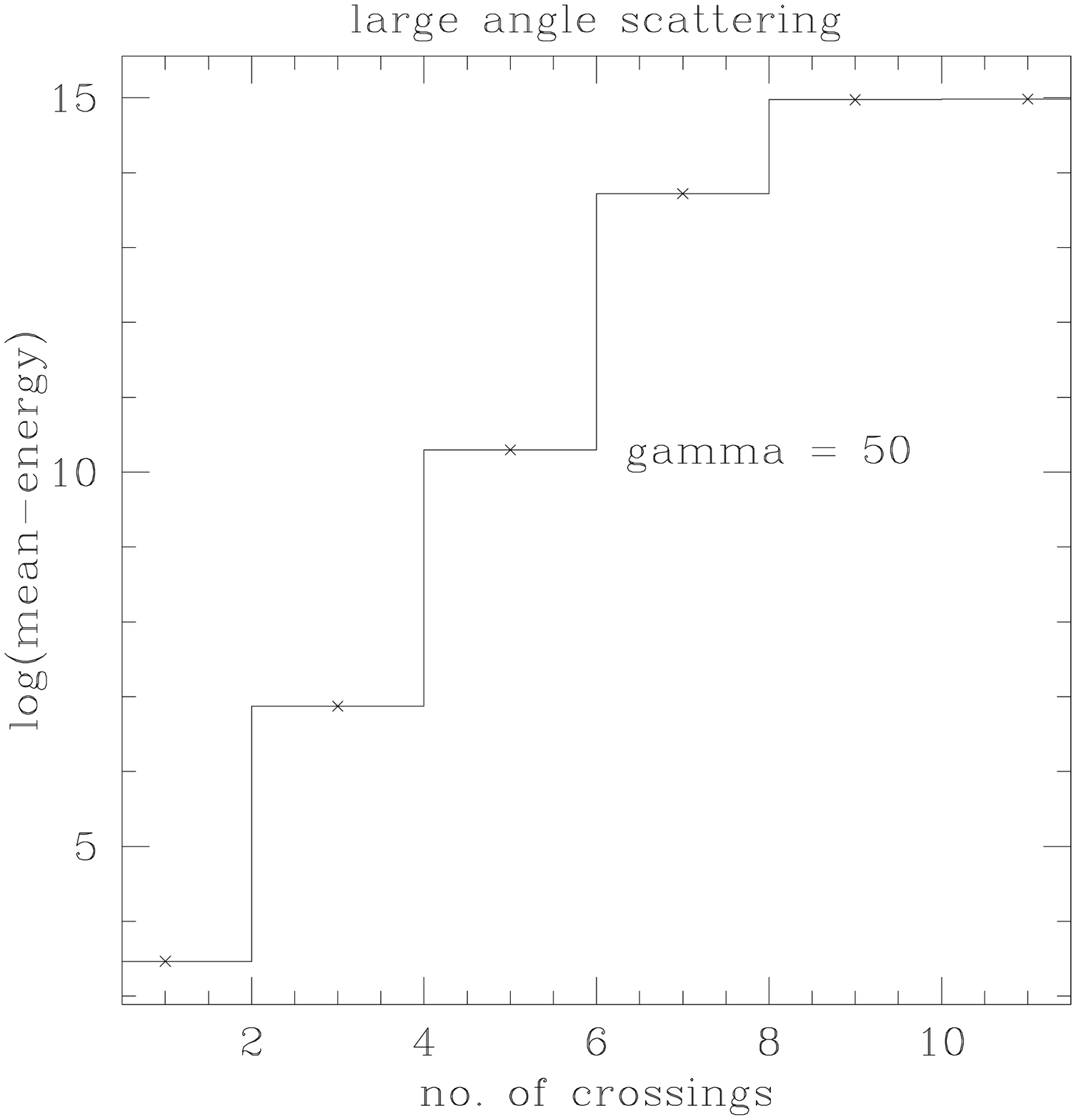, width=5.0cm}
\epsfig{file=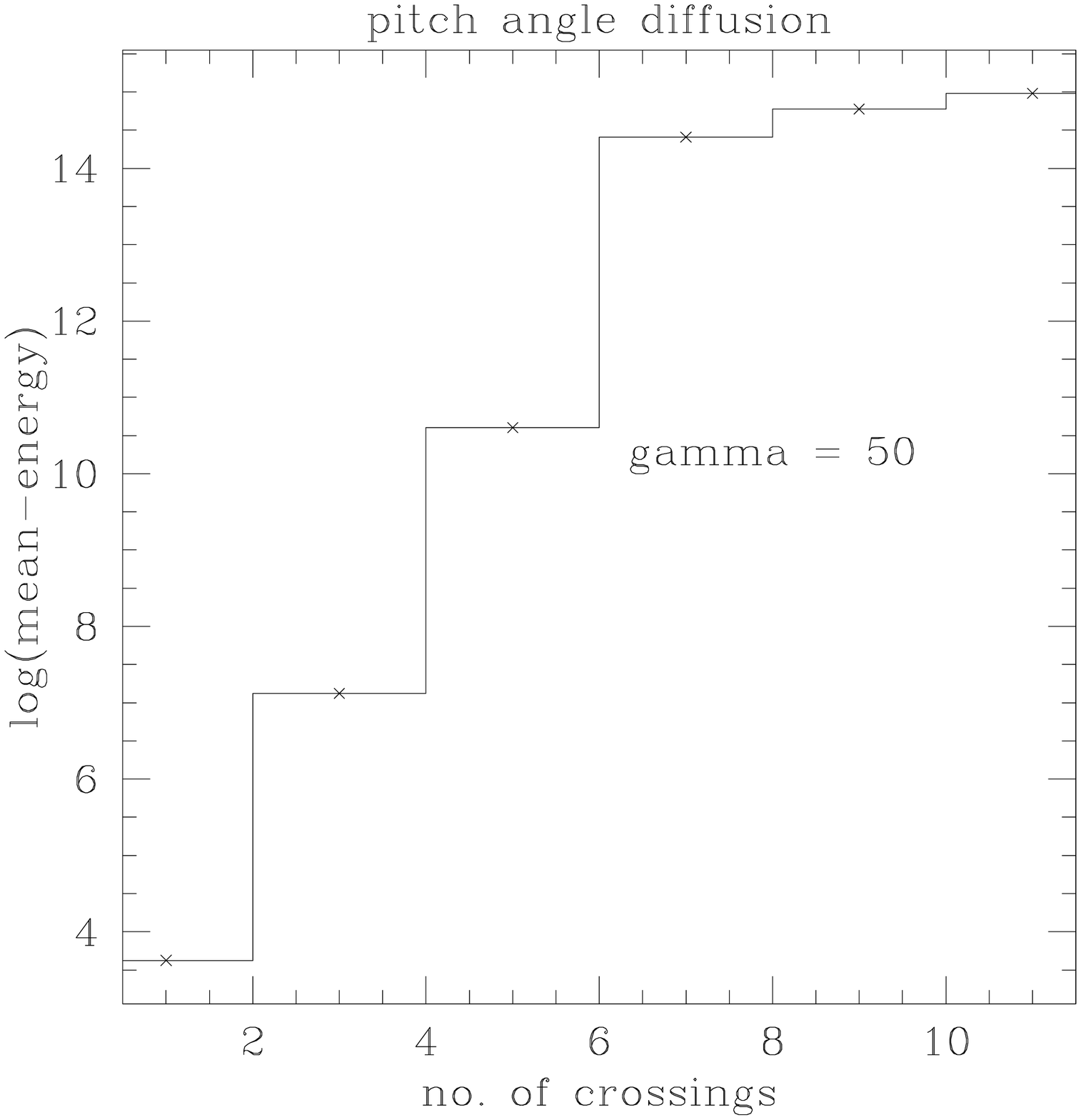, width=5.0cm}
\caption{The logarithm of the mean-energy gain of particles versus the no. of shock 
crossings (1-3-5-7-9-11), where upstream $\Gamma$ equals 50. The energy is measured immediately after
the particle has crossed the shock front. Large angle scattering (left), pitch angle diffusion (right).}
\end{center}
\end{figure}

Particle energy is specified in units of $\gamma$ and hence applies both to electron and ion acceleration.
Figures 2 and 3 
show the logarithm of the particle's mean-energy against the number of
shock crossings for the two scattering models, at $\Gamma$=50 and 990. Here crossings 1 to 3 
(downstream $\rightarrow$ upstream$\rightarrow$ downstream) represent
one cycle.\\ 
The energy is measured immediately after the particle has crossed the
shock front in the shock frame.
In all cases, the energy gain on the first complete cycle, specified by crossing numbers 1 and 3
is $\sim \Gamma^{2}$ but subsequent crossings show a reduced gain, passing through a region 
where the increase is ($\sim\Gamma$) and tending towards the factor $\sim 2$ predicted by
Gallant and Achterberg (1999).

\begin{figure}[t!]
\begin{center}
\epsfig{file=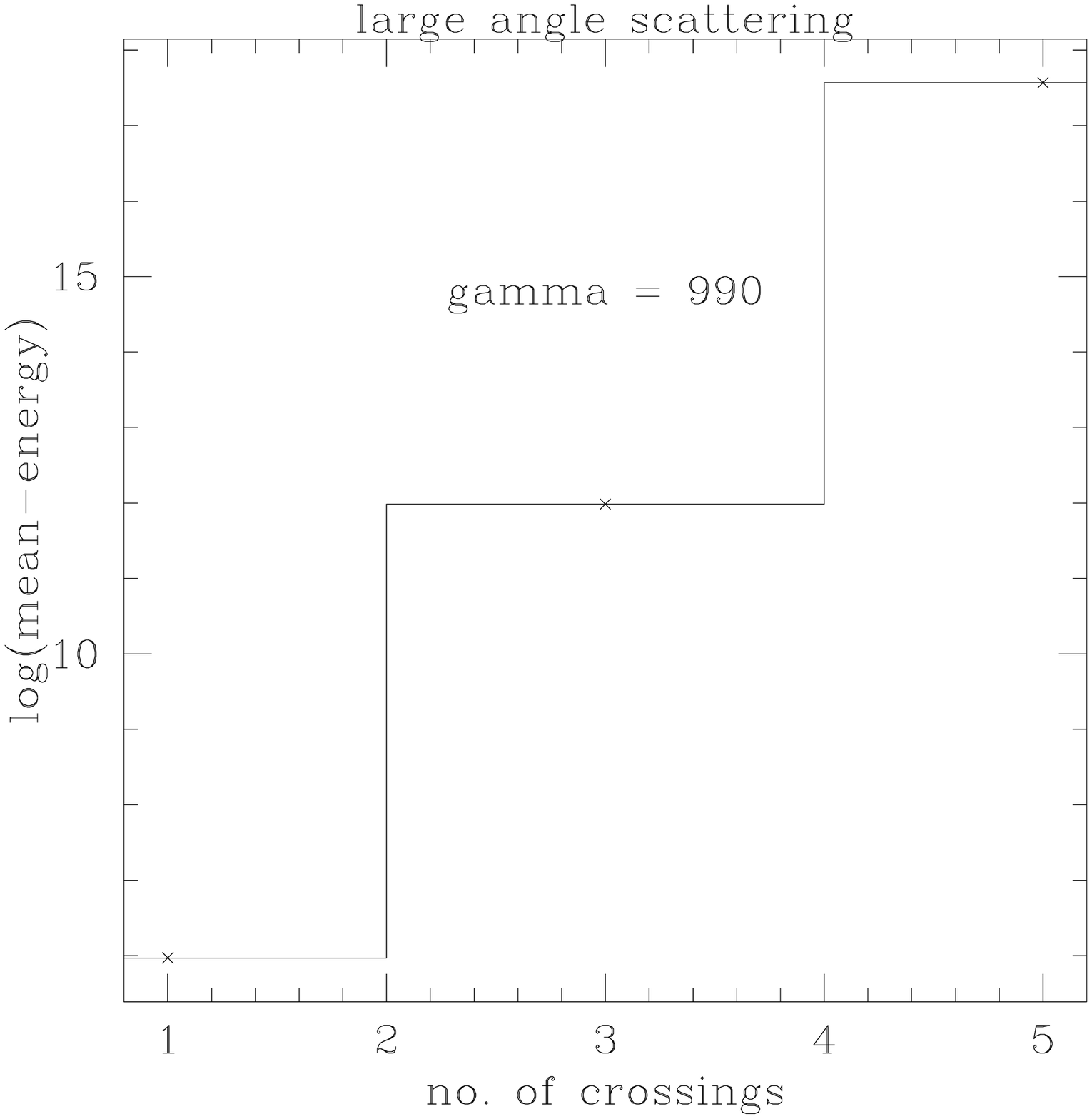, width=5.0cm}
\epsfig{file=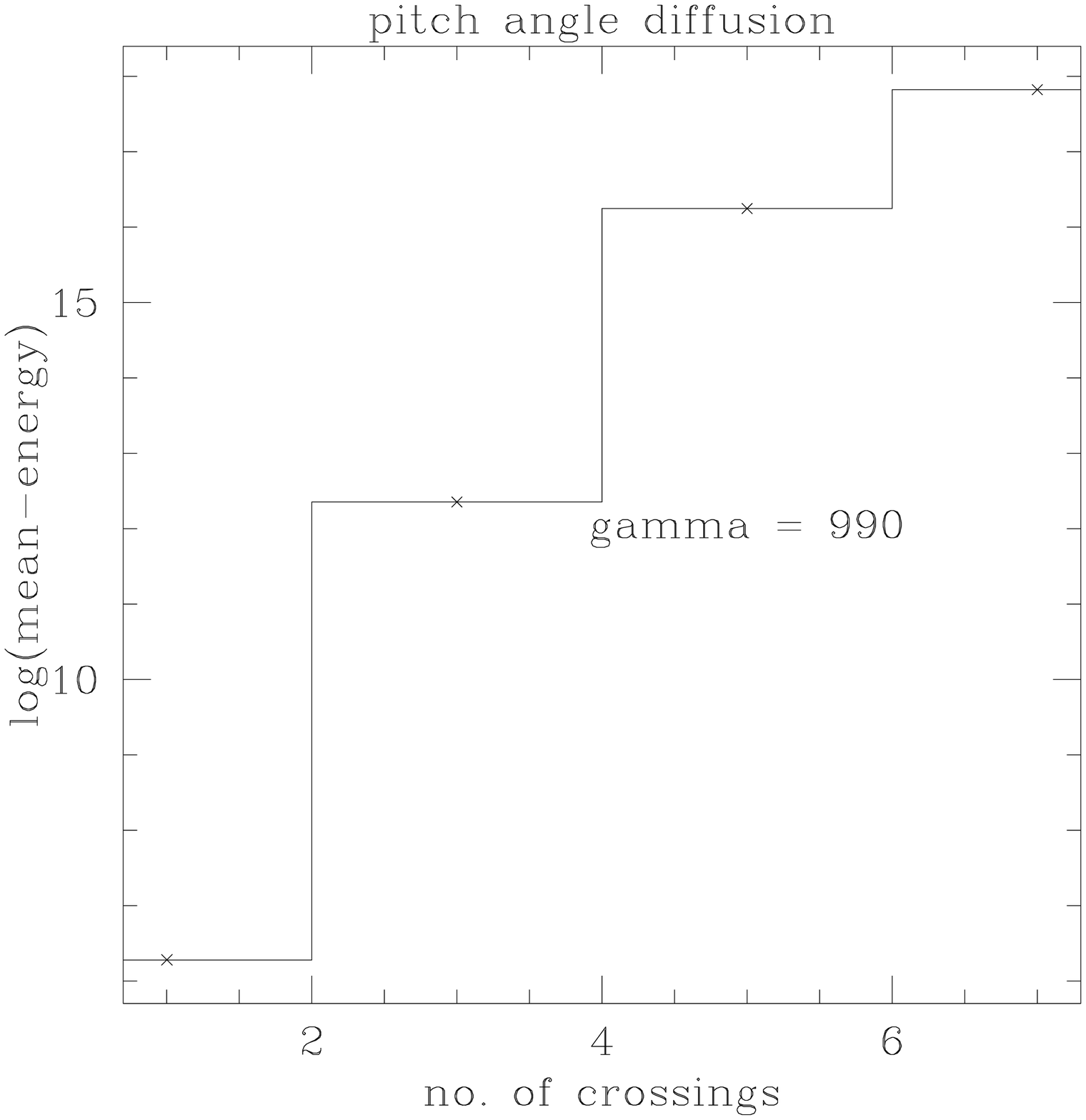, width=5.0cm}
\caption{The logarithm of the  mean energy gain of particles versus the no. of shock 
crossings, employing an upstream flow gamma factor $\Gamma$=990. The energy is measured immediately after
the particle has crossed the shock front. We observe a 
$\Gamma^{2}$ energy boost in the first to the third crossing and significant 
energy multiplication in all subsequent crossings. Large angle scattering (left), pitch angle 
diffusion (right).}
\end{center}
\end{figure}

However, the gain reduction is slower than expected by 
these previous authors and the qualitative idea expressed above. The discrepancy could lie in 
details of the scattering model in that the computation
assumes negligible scattering before the first scatter centre is reached, 
equivalent to almost undeviated guiding centre motion along a field line, whereas the 
Gallant and Achterberg (1999) approach takes the opposite extreme of continued 
deflection in a plane at a large angle to a uniform field.\\
In figure 4 the angular distribution is measured in the shock frame just downstream for
the upstream to downstream transition and just upstream for the down to up transition, for large
angle scattering and upstream flow $\Gamma=200$. Similar plots for small angle scattering
(pitch angle diffusion case) are also presented. We only show the contribution to the distribution
function of particles which have just crossed, not the complete time averaged distribution.\\
Similar results were obtained for $\Gamma=10-990$.
It is evident that there is extreme peaking in the angular distribution towards smaller pitch angles 
in all but one case, continuing the trend noticed by Quenby and Lieu (1989) and Ellison et al. (1990) 
at lower flow $\Gamma$, who found that as the velocities become more relativistic the anisotropies in 
pitch angle become greater.
If the distribution function at the shock is isotropic, the number of particles
between $\mu$ and $\mu+\Delta\mu$  is proportional to $\mu$, taking into account solid angle weighting.

Both scattering cases reveal highly peaked distributions pointing downstream on up to down transition,
due to the beaming of the upstream distribution function as seen by the shock.
\begin{figure}[t!]
\begin{center}
\epsfig{file=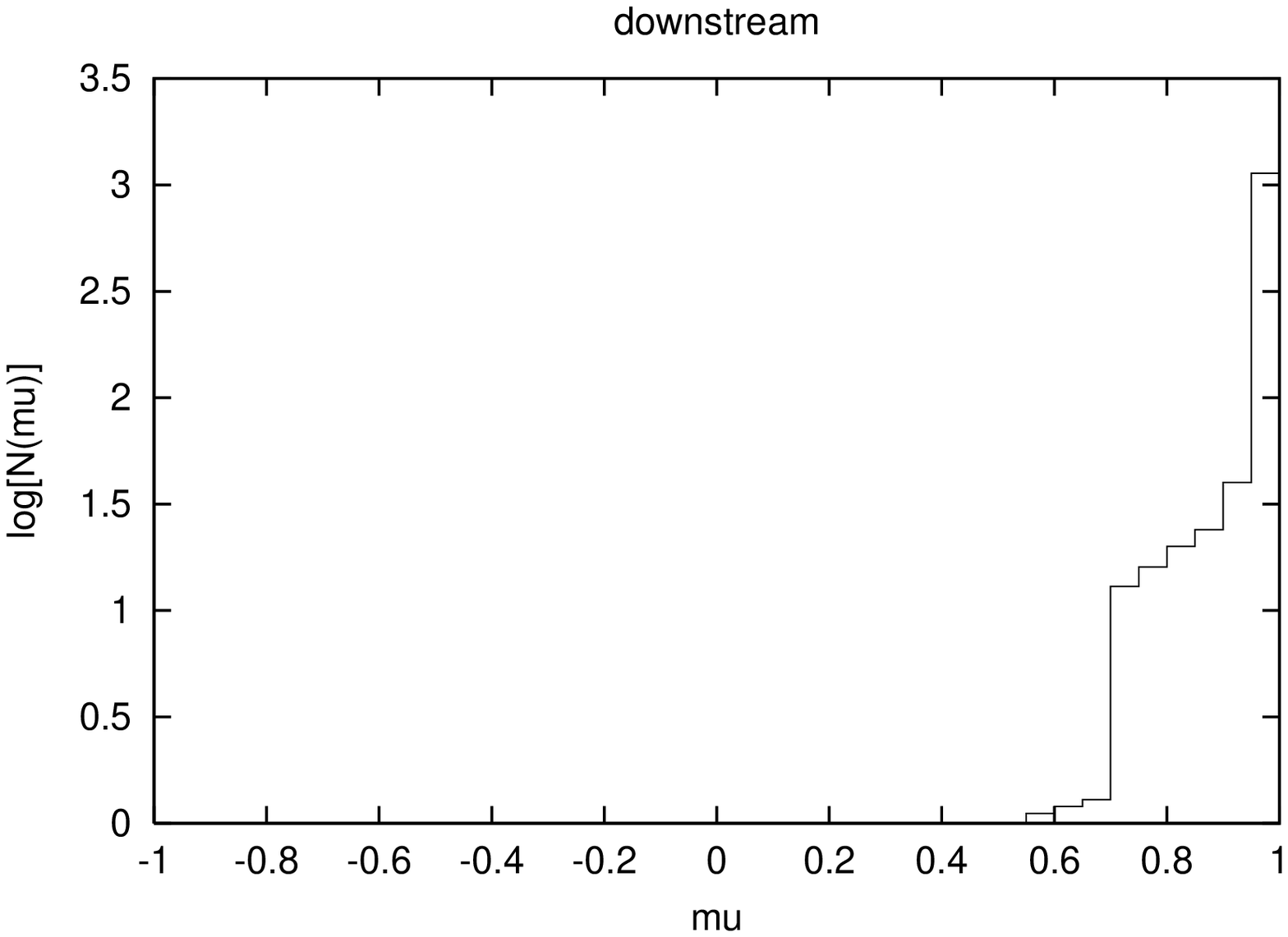, width=6.8cm}
\epsfig{file=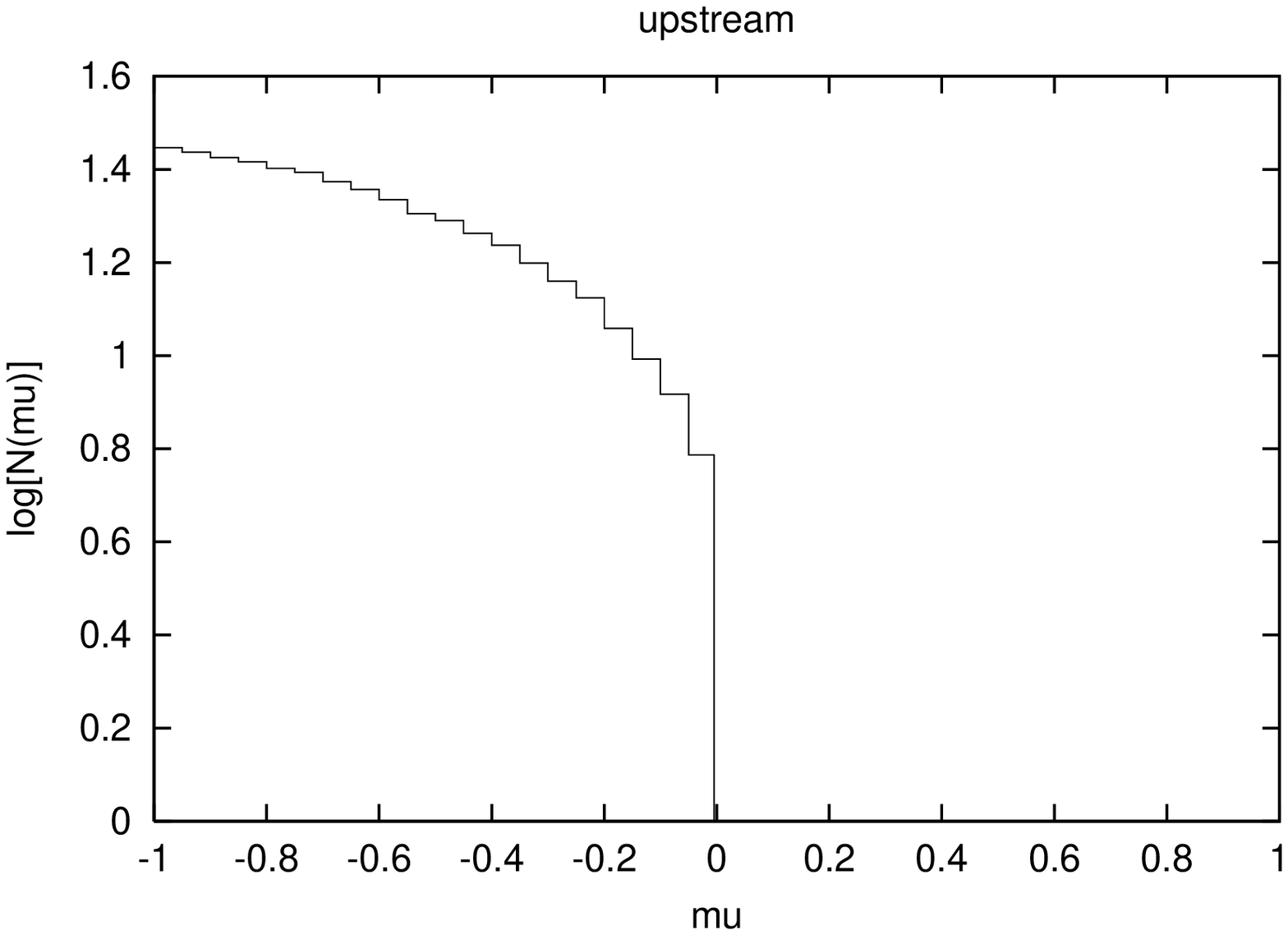, width=6.8cm}
\end{center}
\end{figure}
\begin{figure}[h!]
\begin{center}
\epsfig{file=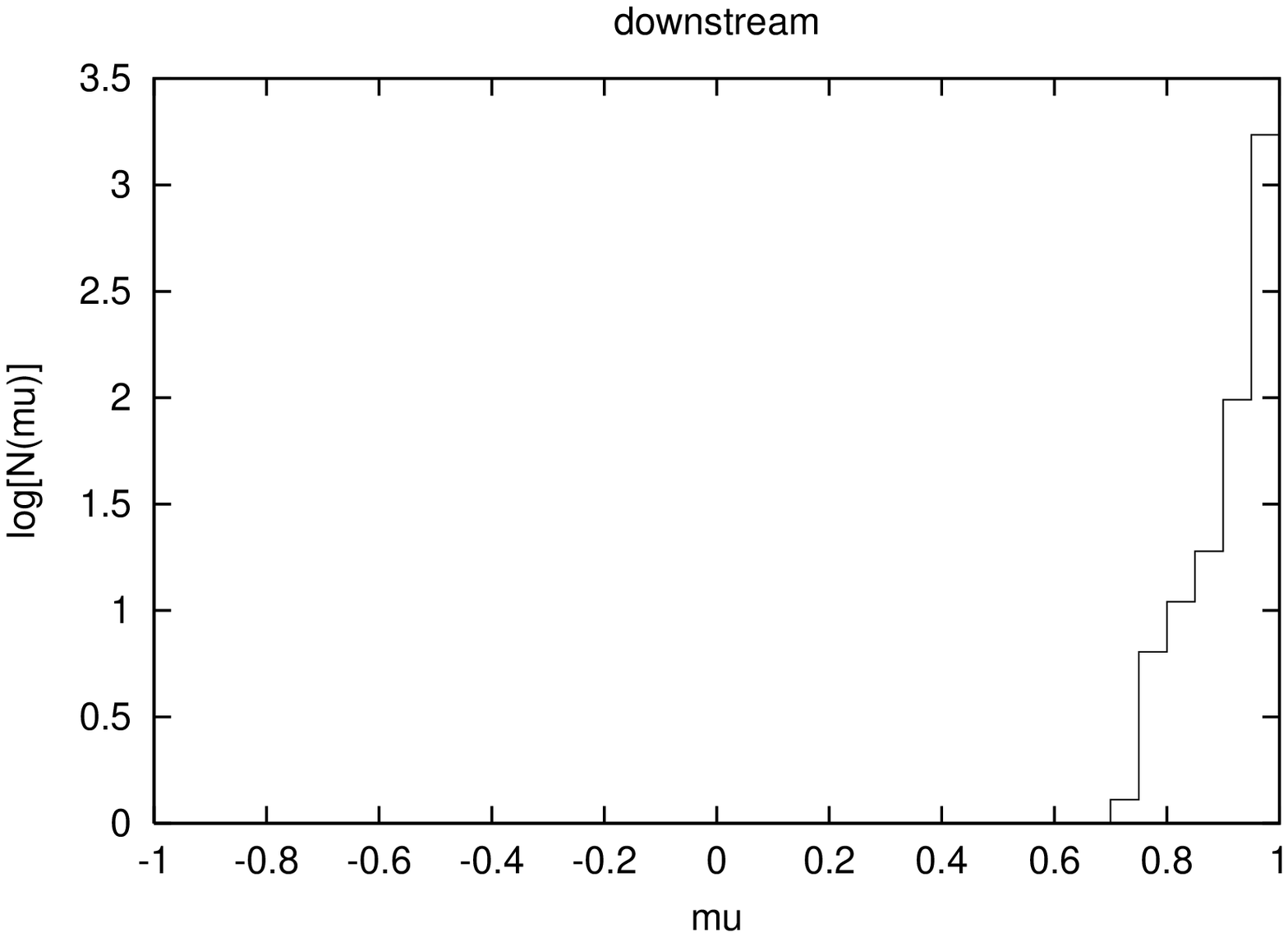, width=6.8cm}
\epsfig{file=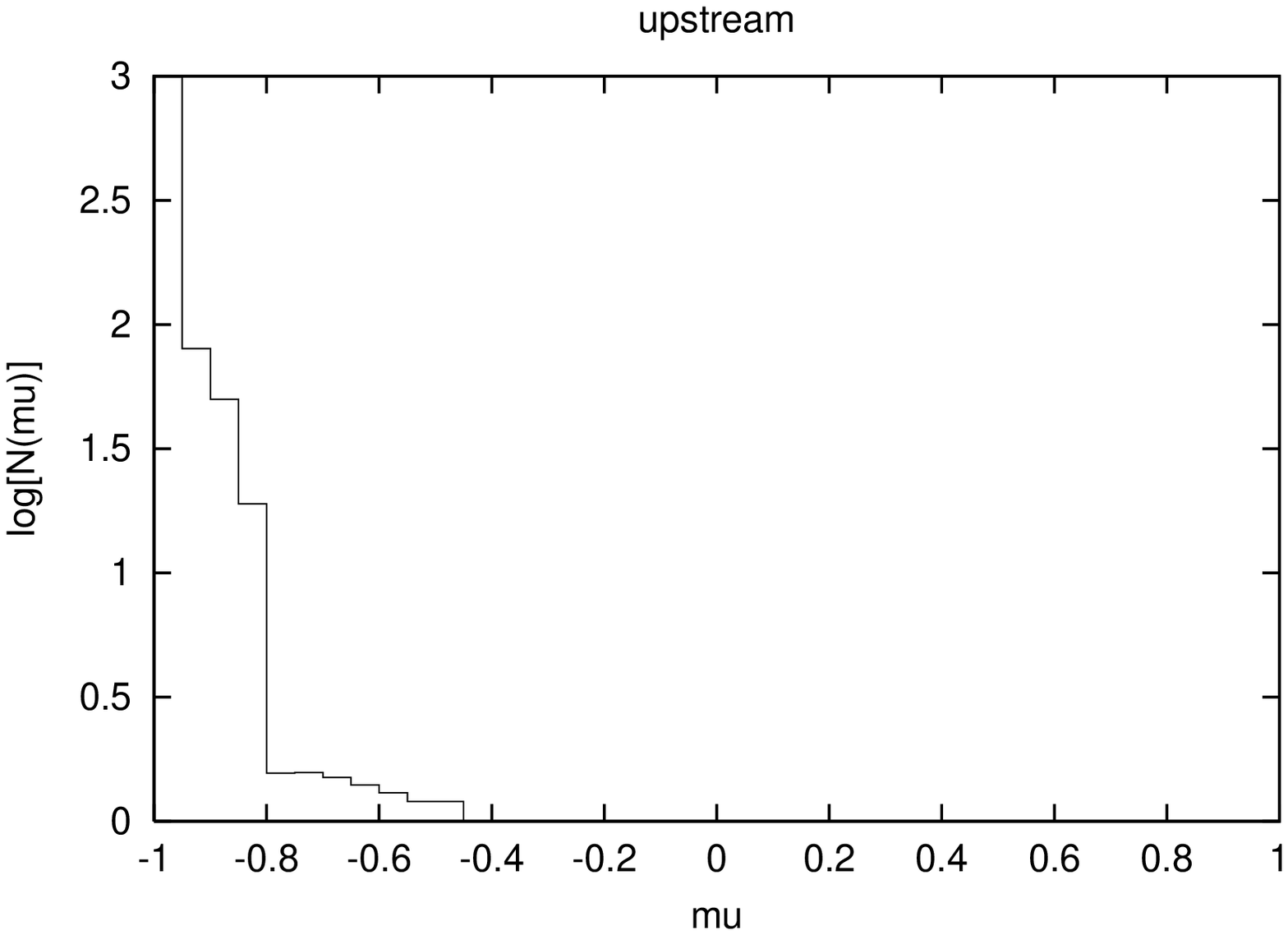, width=6.8cm}
\caption{Angular distribution (mu = $\mu=cos\theta$) in the shock frame 
for $\Gamma=200$, after the particles have crossed the shock from upstream 
$\rightarrow$ downstream (left) and downstream $\rightarrow$ upstream (right). Top plots for 
large angle scattering, bottom plots for pitch angle diffusion.}
\end{center}
\end{figure}
 Downstream to up for large angle scattering produces near triangular distributions 
in the upstream  directed half space, due to the very efficient 
scattering in the model. For small angle scattering, highly peaked distributions are found
due to the lack of time for significant deflection from nearly zero pitch angle before downstream 
re-entry and presumably for those particles which can re-enter upstream.\\
\begin{figure}[t]
\begin{center}
\epsfig{file=plot.lgnew2.g5, width=6.8cm}
\epsfig{file=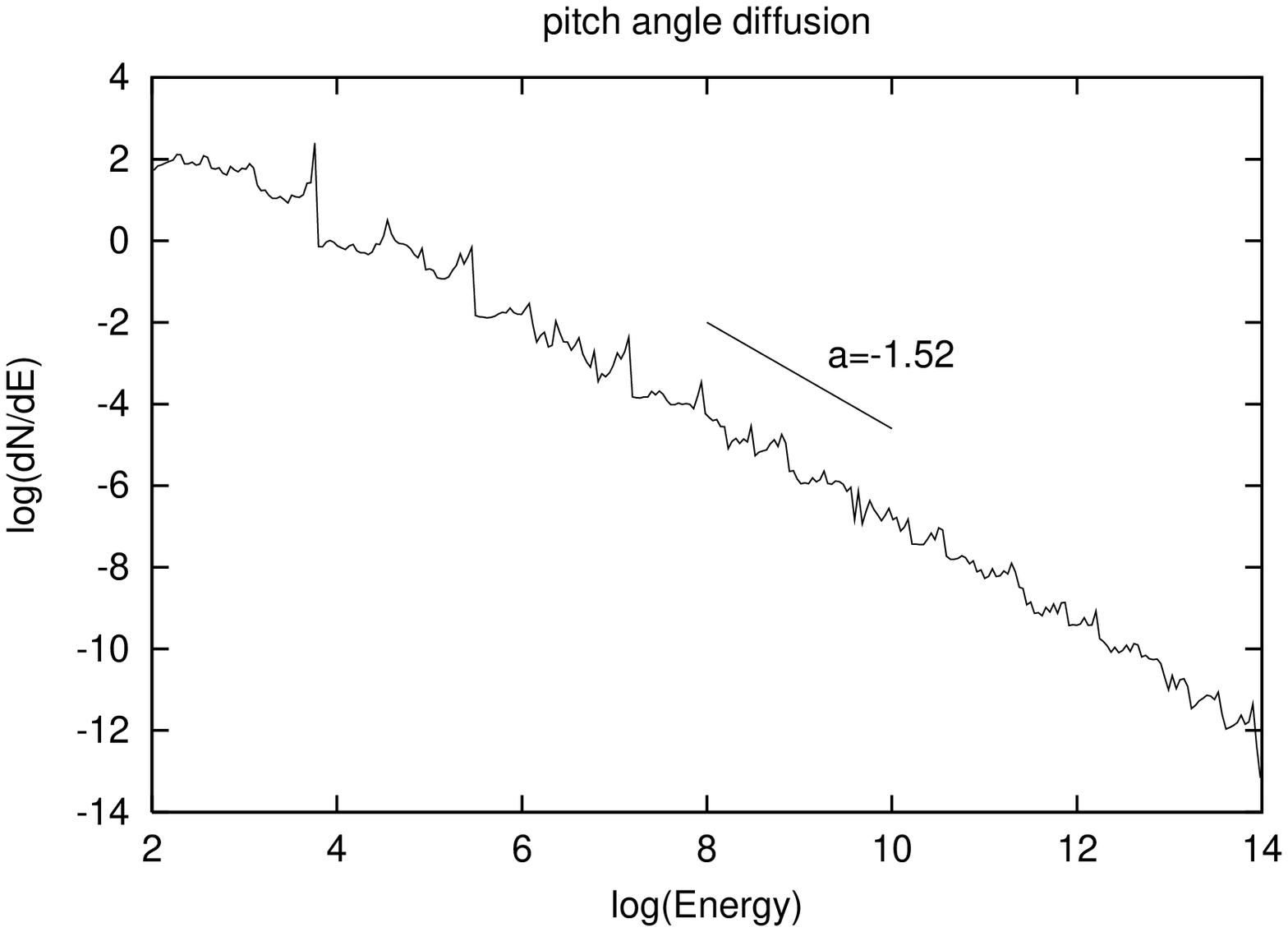, width=6.8cm}
\caption{Spectral shape for upstream $\Gamma$=5. Large angle scattering (left) 
pitch angle diffusion (right). The line shows the mean slope of the spectrum and the number 
is the spectral index found. The smoothness (compared  to higher gamma flows used) and the spectral 
index (especially for the large angle scattering) are consistent with similar work 
presented by Baring (1999).}
\end{center}
\end{figure}
\begin{figure}[t]
\begin{center}
\epsfig{file=plot.lgnew.g50, width=6.8cm}
\epsfig{file=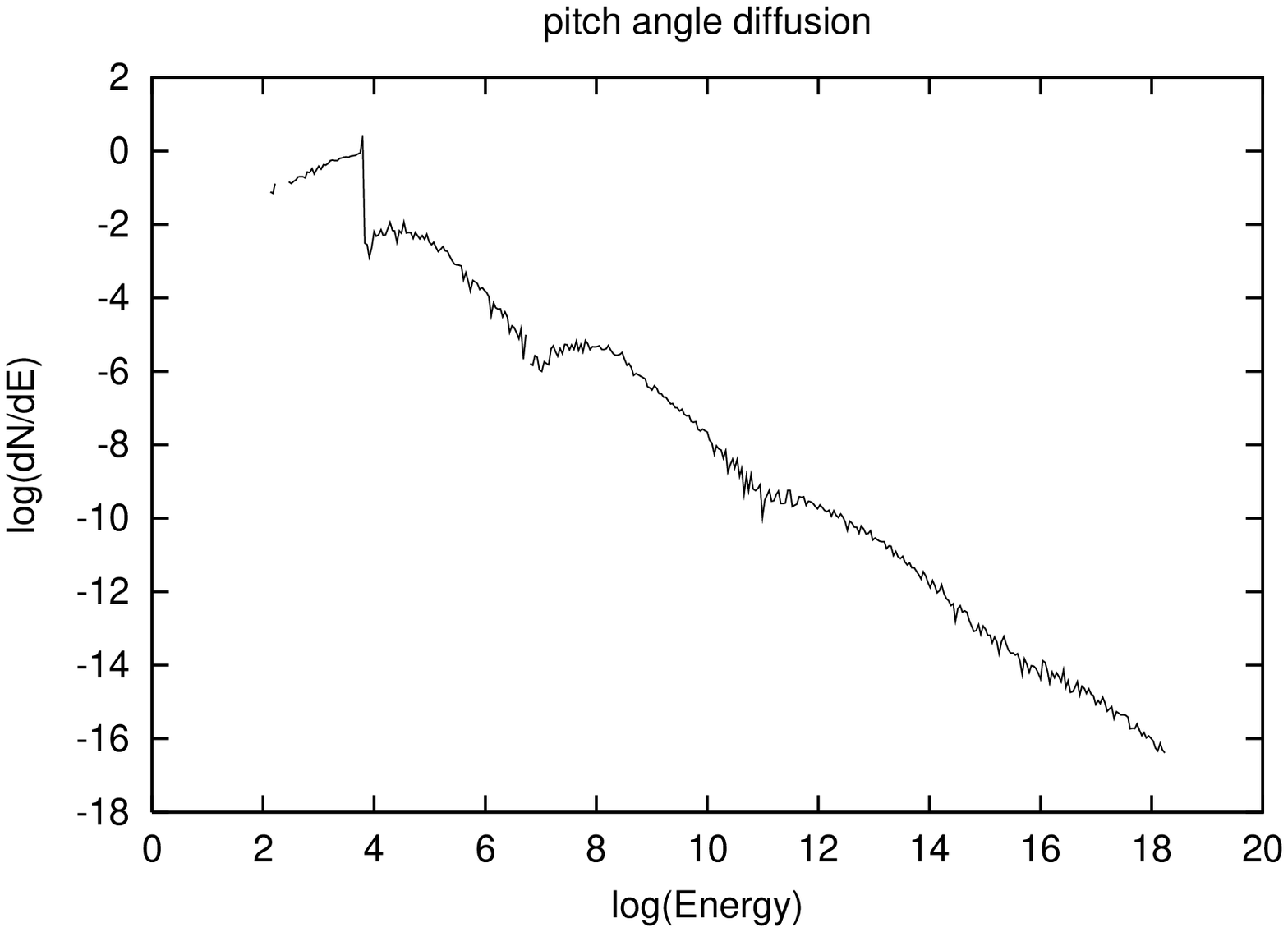, width=6.8cm}
\caption{Left, the spectral shape, for an upstream $\Gamma$=50, 
for large angle scattering. Right, the spectral shape for the same gamma, 
for pitch angle diffusion.}
\end{center}
\end{figure}
\begin{figure}[t]
\begin{center}
\epsfig{file=plot.lgnew.g990, width=6.8cm}
\epsfig{file=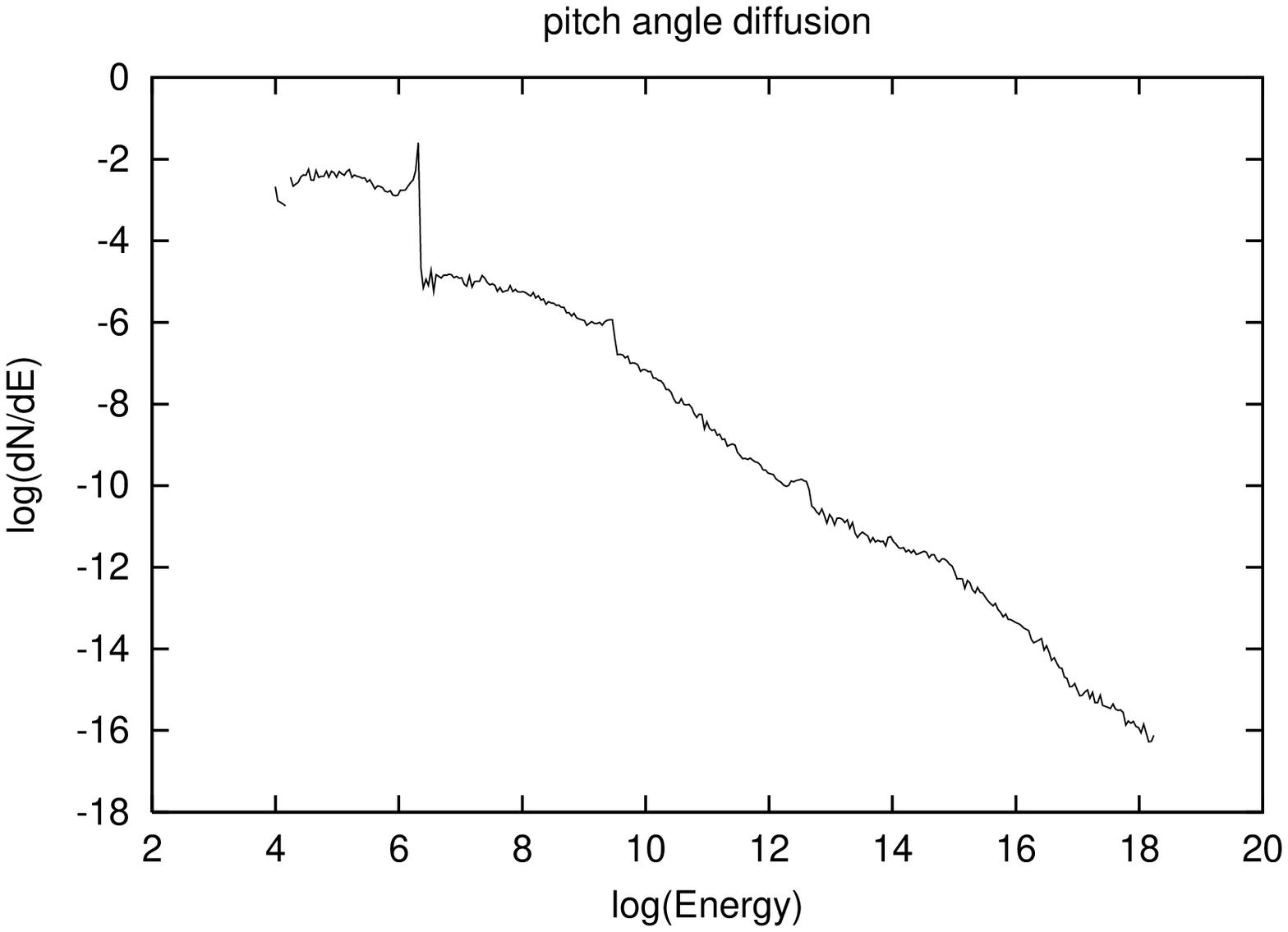, width=6.8cm}
\caption{Left, the spectral shape for an upstream $\Gamma$=990,  for large angle scattering. 
Right, the spectral shape for an upstream  $\Gamma$=990, for pitch angle diffusion.}
\end{center}
\end{figure}
In figures 5,6 and 7 we show the spectral shapes obtained at the shock just downstream,
for large and small angle scattering (pitch angle diffusion) for upstream flow $\Gamma$ 
values of 5, 50 and 990 respectively. We investigated the sensitivity of these spectral shapes
for small angle scattering to downstream boundary condition, and found that the shapes did not change 
significantly for the boundary distance, $r_{b}\geq 2 \cdot 10^{3}\lambda$.
Probably the small chance of return to the shock except for a relatively small subset 
of downstream pitch angle particle 'histories', produced this result and the anisotropy 
in figure 4 is seen for these returning particles.
Note the relatively smooth spectra for the relativistic flow with $\Gamma=5$ (fig. 5)
which becomes plateau-like at the more extreme value of
$\Gamma=990$ (fig. 7) where the effects of individual acceleration cycles are
clearly evident in the large angle case, but not for small angle scattering, where the
spectral shape remains rather 'smooth'. Features
can be seen developing in the lower $\Gamma$ simulations of Quenby and Lieu (1989) 
and Ellison et al. (1990) while Protheroe (2001) shows a similar contrast in behaviour,
between large and small angle scattering models up to $\Gamma=20$.
On any individual cycle, greater energy gain is available for the large angle case because of the 
greater upstream scattering, as demonstrated in the simple analytical approach mentioned above.\\
Hence, the plateaus develop 
for the large angle case with flatter segments, than the smoother pitch angle case spectrum.
An opposing effect, however, is that because the accelerated particles re-entering 
downstream have pitch angles more favourable to eventual loss downstream, the drop between 
plateaus is greater and so the overall spectral slopes in the two cases are rather similar.
The actual mean slope of the differential number spectrum is well above -2, as illustrated by the 
lines of slope -1.45 and -1.52 shown for $\Gamma$=5. This relativistic 
flattening effect is consistent with previous studies by Baring (1999) while, the mechanism of 
plateau development as an acceleration cycle effect is implicit in figure 6 of Protheroe (2001).
It is perhaps not surprising to observe a step-like behaviour in the accelerated spectrum, 
clearly related to groups of particles undergoing different numbers of acceleration cycles,
because the simulations also reveal that about $90\%$  of all particles in the simulation box are 
lost downstream each cycle. One might also expect that small angle scattering has the greater 
chance of smoothing the inherent structure in the acceleration process. Clearly, smoothing is 
much more efficient at lower flow $\Gamma$ where the gain per cycle is less.\\
One can infer that at extreme relativistic flow velocities and different scatter models,
it is difficult to assume that the spectral shape of the accelerated particles follows a  
\textit{smooth} power-law shape, as in the classic non-relativistic first order Fermi 
acceleration mechanism, under all circumstances.


\section{Conclusions}

To what extent is first order Fermi acceleration universally applicable in shocked, turbulent, 
MHD flows ?
Aspects of the problem still needing elucidation include the spectrum produced, the
spectral index, the angular distribution of the particles and non-linear effects in
relativistic flows.\\
We have presented a detailed Monte Carlo approach
for the highly relativistic diffusive acceleration in parallel shocks under the test particle 
approximation, extending previous studies to more extreme values of the upstream flow $\Gamma$ 
and carefully distinguishing the effects of different assumptions concerning the particle scattering.
Comparison is made with results under non-relativistic conditions. 
It is found that the spectral shape in the very relativistic flow regime depends on the
scatter model (large angle or pitch angle diffusion/small angle scattering), with the former model
exhibiting a structure in the spectrum at the shock related to the cycle of acceleration, 
while the later one showing a smoother shape. 
One general indication which supports the validity of  our computational
approach is that the spectral index becomes flatter as the flow becomes
more relativistic, a result to be compared with the conclusions of
Kirk and Schneider (1987a,b).\\
Extreme anisotropy in the shock frame for particles emerging downstream irrespective of
scattering model is observed, due to the beaming of the transmitted flux. High anisotropy
also occurs in the flux emerging upstream in the small angle scattering case, due to the lack of 
time to isotropize the particles but in the large angle scattering model there is a large enough 
probability of scatter  before shock re-entry for near isotropy to set in. This means that for any 
given acceleration cycle, the spectrum is much flatter in the large angle case (plateaus)
because the gain per cycle can be greater. However, evidently because of an enhanced loss downstream 
of the more scattered particles of the large angle scattering case,
the average spectral slope for the two mechanisms is similar. \\
Most important, a significant acceleration rate increase ('speed up'
effect) has been found for both large angle 
and pitch angle diffusion.
This must be seen in conjunction with a 'gamma-squared' energy boosting of the particle 
noted in the first shock cycle, which is in accordance with theoretical indications 
(e.g. Gallant  and Achterberg, 1999) and past simulations (e.g. Lieu and Quenby, 1989). 
There is a gain by a significant factor in subsequent crossings which is $\sim \Gamma$ or even more 
for several cycles. Regarding the acceleration time decrease, computations show that the comparison 
of the simulation (experimental) acceleration time with the analytic non-relativistic time 
constant, indicates a decrease of about 5 with increasing flow gammas (10-$10^{3}$). 
This speed-up is clearly directly related to the boosting of the energy gain.\\
In the theory of GRB production (Meszaros and  Rees, 1993)  
by cosmic fireballs, the $\Gamma \sim 10^{3}$  flows are expected to produce 
some shock acceleration of electrons and protons. Vietri (1995) suggests protons of 
$\gamma \leq 10^{11}$ produce photons of 300 GeV by proton synchrotron radiation and 
neutrinos of $10^{19}$ eV via photopion production and subsequent cascading. This requires only 
two shock acceleration cycles at flow $\Gamma\sim 10^{3}$, a suggestion well supported by the 
results obtained here for both scattering models. 
However, at these high $\Gamma$ values, extreme 'billard ball' scattering is required over distances
much less than $\lambda$ in the large angle case. What finally 
limits the proton maximum energy is escape from the shock turbulence region, as the Larmor 
radius of the accelerated particles increases in size. Despite the more favourable radiative process, 
electrons are not able to produce gamma-ray synchrotron energies higher than those due to these 
protons because the competing synchrotron energy loss severely limits the diffusive shock gain 
(e.g. Gallant et al., 2000) so, the net result is a similar 
upper photon cutoff. Because the accelerated particle spectra at the shock as  
here are not necessarily smooth, it might be expected that a detailed experimental study of 
radiation from this region of a GRB may reveal a related structure. In turn, such a study may 
reveal new insight into the magnetic field regime of the GRB environment.\\ 
The suggestion by Vietri (1995) that the efficient production in GRB of $10^{20}$ eV protons 
constitutes a cosmic ray source, at least in previous epochs, is also supported. This argument 
depends on the similarity of the burst gamma ray energy output and very high energy cosmic ray 
energy flux (Vietri, 1995; Waxman, 1995).\\  
Finally, we may conclude that first order Fermi acceleration in the sense of additive energy 
change over a number of cycles of magnetic discontinuity crossing, appears to be applicable 
to $\Gamma\sim 10^{3}$ flows where, there is no test particle reflection at the discontinuity. However, 
the large gradients in particle distribution  function appearing in the simulations argue against 
a spatial diffusion approximation description of the phenomenon.\\

{\noindent \bf Acknowledgments}\\

We wish to thank Prof. Drury and the unknown referee for their valuable comments and 
suggestions regarding this work.


\end{document}